\title{The {\sc eagle} simulations of galaxy formation: \\ Public release of particle data}
\author{The EAGLE team}
\date{\today}

\documentclass[10pt, a4paper]{article}

\usepackage{graphicx}
\usepackage{amsmath}
\usepackage{amsfonts}
\usepackage{amssymb}
\usepackage{graphicx}
\usepackage{xspace}
\usepackage{aas_macros}
\usepackage{hyperref}
\usepackage{array}
\usepackage{multirow}
\usepackage{placeins}
\usepackage{color}

\usepackage{geometry}
\DeclareFixedFont{\ttb}{T1}{txtt}{bx}{n}{8} 
\DeclareFixedFont{\ttm}{T1}{txtt}{m}{n}{8}  
\usepackage{color}
\definecolor{deepblue}{rgb}{0,0,0.5}
\definecolor{deepred}{rgb}{0.6,0,0}
\definecolor{deepgreen}{rgb}{0,0.5,0}
\definecolor{grey}{rgb}{0.4,0.4,0.4}
\usepackage{listings}
\newcommand\pythonstyle{\lstset{
language=Python,
basicstyle=\ttm,
otherkeywords={self},             
keywordstyle=\ttb\color{deepblue},
emph={MyClass,__init__},          
emphstyle=\ttb\color{deepred},    
stringstyle=\color{deepgreen},
frame=tb,                         
showstringspaces=false            %
}}
\newcommand\pythonexternal[2][]{{
\pythonstyle
\lstinputlisting[#1]{#2}}}

\newcommand{\groupnumber}{Friends of Friends (FoF) group number this particle
belongs to in this snapshot. Values range from 1-$N$, where $N$ is the total
number of FoF groups. Values of $2^{30}$ indicate this particle does not belong
to any group.}
\newcommand{\subgroupnumber}{Subgroup number (as defined by {\sc subfind}) this
particle belongs to. Values range from 0-($N$-1) where $N$ is the total number
of subgroups for this particular FoF group. Values of $2^{30}$ indicate this
particle does not belong to any subgroup. Subgroup number 0 refers to the
central subgroup, subgroup numbers greater than 0 refer to satellites.}
\newcommand{\velocity}{The peculiar velocity, $a{\rm d}{\bf x}/{\rm d}t$ (see
the Appendix of \cite{mcalpine2016} for more details).}

\newcommand{\coordinates}{Co-moving coordinates.}
\newcommand{\eagle}{{\sc eagle}}
\newcommand{\virgo}{{\sc virgo}}
\newcommand{\REF}{{\sc reference}}
\newcommand{\Msun}{{\hbox{${\rm M}_\odot$}}}
\newcommand{\hdf}{{\sc hdf5}}
\newcommand{\gadget}{{\sc gadget}}
\begin{document}
\maketitle

\begin{abstract}
This manual accompanies the release of the particle data for 24 simulations of
the \eagle\ suite of cosmological hydrodynamical simulations of galaxy
formation by the \virgo\ consortium. It describes how to download these
snapshots and how to extract datasets from them, emphasising the meaning of
variables, and their units. We provide examples for extracting the particle
data in {\sc python}. This data release complements our earlier release of
numerous integrated properties of the galaxies in \eagle\ through an {\sc sql}
relational database. This database has been updated to include the additional
simulations that are part of the present data release. Scientists wanting to
use \eagle\ may find it useful to first investigate whether their analysis can
be performed using the database, before accessing the particle data.  The
particles in the snapshot files are indexed by a peano-hilbert key. This allows
for an eased extraction of simply connected spatial volumes, without needing to
read the entire snapshot. This makes it possible to analyse many aspects of
galaxies using modest computing resources, even when using \eagle\ simulations
with large numbers of particles. A reading routine is provided to simplify this
process.
\end{abstract}

\section{Introduction}
\subsection{The \eagle\ simulations}
The Virgo consortium's {\bf E}volution and {\bf A}ssembly of {\bf G}a{\bf
L}axies and their {\bf E}nvironments (hereafter {\sc eagle}) project consists
of a suite of cosmological hydrodynamical simulations designed to enable the
study of the formation and evolution of a population of galaxies in a
cosmologically representative volume \cite{schaye2015,crain2015}, adopting the
cosmological parameters advocated by the Planck Collaboration (\cite{Planck13})
($\Omega_{\rm m} = 0.307$, $\Omega_\Lambda = 0.693$, $\Omega_{\rm b} =
0.04825$, $h = 0.6777$, $\sigma_8 = 0.8288$, $n_{\rm s} = 0.9611$, $Y=0.248$). 
	
All simulations were performed with the {\sc gadget-3} tree-SPH code, which is
based on the {\sc gadget-2} code described by \cite{springel2005}, but
extensively modified as described in detail by \cite{schaye2015} and references
therein.  Modifications include changes to the hydrodynamics and time-stepping
schemes referred to as {\sc anarchy}, and the implementation of a large number
of \lq subgrid\rq\ modules that account for physical processes below the
resolution scale (radiative cooling and heating in the presence of an imposed
optically thin radiation background, star formation, stellar evolution, metal
enrichment, feedback from stars, seeding and growth by accretion and merging of
supermassive black holes and feedback from accreting black holes). These
subgrid modules are described by \cite{schaye2015} and references therein.
Numerical parameters for the \eagle\ \REF\ model associated with the subgrid
modules were calibrated to a limited subset of $z=0$ observations of galaxies,
namely: the galaxy stellar mass function, the sizes of galaxies, and the black
hole mass - stellar mass relation. The motivation for doing so is detailed by
\cite{schaye2015}, with the calibration strategy described in detail by
\cite{crain2015}.
	
Simulations were performed in cubic volumes with lengths of $L=12$, 25, 50 and
100 co-moving mega-parsec (cMpc) on a side, at a range of resolutions.
Simulations with an initial baryonic particle mass (SPH mass) of $m_{\rm
g}=2.26\times 10^5\,{\rm M}_\odot$ are referred to as \lq high resolution\rq\
and those with an initial baryonic particle mass of $m_{\rm g}=1.81\times
10^6\,{\rm M}_\odot$ referred to as \lq intermediate resolution\rq. The
corresponding (\lq Plummer equivalent\rq) maximum gravitational softening
lengths are $\epsilon_{\rm prop}=0.35$~pkpc (proper kilo-parsec) and 0.70~pkpc
respectively, with the code switching to a softening that is a constant,
$\epsilon_{\rm com}$, in co-moving coordinates at $z \geq 2.8$.
Table~\ref{table:sims} summarises these properties for key runs, and
illustrates the naming convention, for example L0025N0376 refers to a simulation
with $L=25$ cMpc that starts with $2\times 376^3$ (dark matter and gas)
particles.

In addition to the \REF\ simulations, the \eagle\ suite comprises runs where
one or more of the subgrid modules, or parameters of these modules, were
changed.  The {\sc recal} model, {\sc recal}L0025N0752, is calibrated to the
same $z=0$ galaxy properties as \REF, with the (relatively small) changes to
the subgrid parameters as a consequence of the higher resolution compared to the
default \lq intermediate resolution\rq\ runs. Other models, e.g, {\sc
WeakFB}L0025N0376 or {\sc StrongFB}L0025N0376, are not re-calibrated. These
variations to subgrid parameters are aimed at understanding the effect of
changing parameters one at a time (in the example, the strength of feedback
from star formation). The models are summarised in Table~\ref{table:runs}, with
a short description of how they differ from \REF.

\eagle\ has generated several spin off projects that use the same (or very
similar) simulation code and models. These include zooms of Local Group-like
regions (the {\sc apostle project} described by \cite{sawala2016,fattahi2016});
zooms of galaxy clusters (the {\sc C-eagle} and {\sc hydrangea} projects
described by \cite{bahe2017, barnes2017}); simulations with warm dark matter
\cite{lovell2016}; and zooms of Milky Way-like galaxies that include
non-equilibrium chemistry \cite{oppenheimer2016}. Currently, the data from
those spin off projects is not part of this data release. The {\sc Virgo}
consortium releases these particle data in the hope that they will be useful to
the community. We are keen to receive feedback and suggestions concerning how
this release could be made more useful.

\begin{table}
\caption{Box sizes and resolutions of the main \eagle\ simulations.  From
left-to-right the columns show: simulation name suffix; comoving box size;
number of dark matter particles (there is initially an equal number of baryonic
particles); initial baryonic particle mass; dark matter particle mass;
comoving Plummer-equivalent gravitational softening length; maximum proper
softening length.} 
\label{table:sims}
 \begin{center}
\begin{tabular}{lrrrrrrr}
\hline
Name & $L$ & $N$ & $m_{\rm g}$ & $m_{\rm dm}$ &
$\epsilon_{\rm com}$ & $\epsilon_{\rm prop}$ \\  
& (cMpc) & & ($\Msun$) & ($\Msun$) & (ckpc) & (pkpc)\\
\hline 
{\sc Ref}L0025N0376 &  25 & $376^3$ & $1.81\times 10^6$ & $9.70\times 10^6$ & 2.66 &0.70\\
{\sc Recal}L0025N0752 &  25 & $752^3$ & $2.26\times 10^5$ & $1.21\times 10^6$ & 1.33 & 0.35\\
{\sc Ref}L0050N0752 &  50 & $752^3$ & $1.81\times 10^6$ & $9.70\times 10^6$ & 2.66 &0.70\\
{\sc Ref}L0100N1504 & 100 & $1504^3$ & $1.81\times 10^6$ & $9.70\times 10^6$ & 2.66 &0.70\\
\hline
\end{tabular}
\end{center}
\end{table}

\begin{table}
\caption{List of available simulations. Simulation names end with
``LxxxxNyyyy'' where xxxx is the side length of the simulation box in comoving
Mpc and yyyy$^3$ is the number of particles (for both baryons and dark matter).
Note that AGNdT9 uses a different subgrid black hole accretion disc viscosity
than C15AGNdT9.}
\label{table:runs}  
\begin{center}
\footnotesize
\begin{tabular}{lll}
\hline
Name & Resolution & Description \\ \hline\hline
\multicolumn{3}{l}{\emph{Small test volume}}\\
\hline
{\sc Ref}L0012N0188 & Intermediate & Reference model\\
\hline
\multicolumn{3}{l}{\emph{Models from \cite{schaye2015}}}\\
\hline
{\sc Ref}L0025N0376 & Intermediate & Reference model\\
{\sc Ref}L0025N0752 & High & Higher resolution version evolved using Reference
parameters\\ {\sc Recal}L0025N0752 & High & Higher resolution version recalibrated to
same data as Reference model\\
{\sc Ref}L0050N0752 & Intermediate & Reference model\\
{\sc AGNdT9}L0050N0752 & Intermediate & Higher AGN heating temperature and lower
subgrid black hole accretion disc viscosity\\
{\sc Ref}L0100N1504 & Intermediate &
Reference model\\ \hline \multicolumn{3}{l}{\emph{Models calibrated to the
$z=0$ galaxy stellar mass function from \cite{crain2015}}}\\
\hline
{\sc FBconst}L0050N0752 & Intermediate &Constant stellar feedback efficiency\\
{\sc FBsigma}L0050N0752 & Intermediate &Stellar feedback dependent on dark matter
velocity dispersion\\
{\sc FBZ}L0050N0752 & Intermediate &Stellar feedback dependent
only on metallicity (not on density)\\
\hline
\multicolumn{3}{l}{\emph{Model variations from \cite{crain2015}}}\\
\hline {\sc eos1}L0025N0376 & Intermediate
&Slope of equation of state imposed on the ISM equals 1.0 (i.e.\ isothermal)\\
{\sc eos53}L0025N0376& Intermediate &Slope of equation of state imposed on the ISM
equals $5/3$ (i.e.\ adiabatic)\\
{\sc FixedSfThresh}L0025N0376 & Intermediate &Star
formation threshold of $n_{\rm H} = 0.1~{\rm cm}^{-3}$, independent of
metallicity\\
{\sc WeakFB}L0025N0376 & Intermediate & Half as much energy feedback
from star formation \\
{\sc StrongFB}L0025N0376 & Intermediate &Twice as much energy
feedback from star formation\\
{\sc ViscLo}L0050N0752 & Intermediate &$10^2$ times
lower subgrid black hole accretion disc viscosity\\
{\sc ViscHi}L0050N0752 &
Intermediate &$10^2$ times higher subgrid black hole accretion disc viscosity\\
{\sc C15AGNdT8}L0050N0752 & Intermediate &$10^{0.5}$ times lower AGN heating
temperature \\
{\sc C15AGNdT9}L0050N0752 & Intermediate & $10^{0.5}$ times higher AGN
heating temperature\\ \hline
\multicolumn{3}{l}{\emph{Models without black holes}}\\
\hline
{\sc NoAGN}L0025N0376 & Intermediate & Same as Reference model, but without black holes\\
{\sc NoAGN}L0050N0752 & Intermediate & Same as Reference model, but without black holes\\
\hline
\multicolumn{3}{l}{\emph{Collisionless simulations with only dark matter
  particles (with mass $\Omega_{\rm m} / (\Omega_{\rm m} - \Omega_{\rm
  b})$ higher than in the reference model)}}\\
\hline
{\sc DMONLY}L0025N0376 & Intermediate\\
{\sc DMONLY}L0025N0752 & High\\
{\sc DMONLY}L0100N1504 & Intermediate\\
\hline

\end{tabular}
\end{center}
\end{table}

\subsection{The \eagle\ database - FoF and {\sc subfind} groups}
Haloes are identified in the simulations using the friends-of-friends (FoF) and
spherical over-density algorithms.  Baryonic particles (gas, stars and black
holes) are assigned to the same halo as the nearest dark matter particle, if
that particle belongs to a halo. Galaxies are identified as self-bound
substructures using the {\sc subfind} algorithm of \cite{springel01,dolag09}.
	
Particles in a FoF halo are tagged with a \texttt{GroupNumber} (particles with
\texttt{GroupNumber}=$2^{30}$ do not belong to any group). This integer runs
from 1 (first group) to $N$ (total number of groups). Particles in a self-bound
substructure are tagged with a \texttt{SubGroupNumber}, which ranges from 0
(the main galaxy in this FoF group), to $N-1$ (where $N$ is the number of
subgroups in this group).  Particles with \texttt{SubGroupNumber}=$2^{30}$ do
not belong to any subgroup.  It is important to realise that
\texttt{GroupNumber} and \texttt{SubGroupNumber} refer to a {\em given}
snapshot: groups with the same value of \texttt{GroupNumber} in {\em different}
snapshots are generally not the same physical structure.

Many properties of galaxies and haloes, such as their masses, positions,
velocities and spins, can be easily accessed using the {\color{blue}{\sc sql}
\href{http://icc.dur.ac.uk/Eagle/database.php}{database}} documented by
\cite{mcalpine2016}. Since the original database release, the galaxy data has
been extended to include mock observables including intrinsic broad-band
colours and colours computed using a dust screen model, as described by
\cite{trayford15}, and broadband colours computed using dust radiative
transfer, as described by \cite{camps16, trayford17}. Images of galaxies are
available as well, and can be downloaded using {\sc sql} queries. 

We recommend the database as a first approach to analysis of the \eagle\ simulations. 

\section{The \eagle\ particle data} \label{sect:eagle_pdata}
\subsection{Downloading the data}
The particle data for the runs described by \cite{schaye2015} and
\cite{crain2015}, summarised in Table~\ref{table:runs}, can be downloaded from
{\color{blue} \url{http://icc.dur.ac.uk/Eagle/database.php}}, after
registration. This document merely serves as a pointer to, and brief
description of, the data and it is not meant as a reference for the \eagle\
simulations (for that see \cite{schaye2015,crain2015}). Below we use the
\texttt{Courier font} to denote snapshot variables.

\subsection{The snapshot format}
\begin{table}
\label{table:file_layout}
\caption{\hdf\ groups in snapshot files}
\begin{center}
\footnotesize
\renewcommand{\arraystretch}{1.5}
\begin{tabular}{>{\ttfamily}p{4cm}p{3cm}p{9.5cm}}
\hline
Name & Type & Description \\ \hline\hline
snap\_028\_z000p000.0.hdf5 & HDF5 FILE & single file of snapshot 28 ($z=0$) \\
\hline
Config & HDF5 GROUP & Configuration parameters for the GADGET code, and svn subversion revision number \\
Constants & HDF5 GROUP & Numerical values for physical constants used in the run \\
HashTable & HDF5 GROUP & Contains information needed to identify particles in hash cells \\
Header & HDF5 GROUP & \gadget\ header, including Hubble parameter $h$ and expansion factor $a$ \\
Parameters & HDF5 GROUP & Initial and solar abundances of all chemical elements tracked \\
PartType0 & HDF5 GROUP & Gas particle data \\
PartType1 & HDF5 GROUP & Dark matter particle data  \\
PartType4 & HDF5 GROUP & Star particle data \\
PartType5 & HDF5 GROUP & Black hole particle data \\
RuntimePars & HDF5 GROUP & Input parameters for this run \\
Units & HDF5 GROUP & Units of length, mass, time, and those derived from it, in cgs units \\
\hline
CGSConversionFactor & HDF5 ATTRIBUTE & Multiplier to convert a variable into cgs units \\
VarDescription & HDF5 ATTRIBUTE & Text description of variable \\
aexp-scale-exponent & HDF5 ATTRIBUTE & Physical quantity is $a^{\rm aexp-scale-exponent}$ times variable \\
h-scale-exponent & HDF5 ATTRIBUTE & Physical quantity is $h^{\rm h-scale-exponent}$ times variable \\
\hline

\end{tabular}
\end{center}
\label{table:groups}
\end{table}

Particle data are output in snapshots - the state of the system at a given
redshift - with different redshifts, $z$, corresponding to different snapshots
(29 snapshots from $z=20$ to $z=0$). Each snapshot is distributed over several
files, and to extract all particles from a given snapshots one must read {\em
all} files - even when reading a single variable such as, for example, the
coordinates of dark matter particles (Section \ref{sect:python_examples}
contains an example in {\sc Python} of how to read snapshot datasets in this
manner). Readers unfamiliar with {\sc gadget} may want to read Volker Springel's
description of the format from the
{\color{blue}\href{https://wwwmpa.mpa-garching.mpg.de/gadget/users-guide.pdf}{\sc
gadget} manual}.

\subsubsection{The {\sc hdf5} format}
Individual snapshot files are written in the binary
{\color{blue}\href{https://www.hdfgroup.org/HDF5/}{\sc hdf5} format}. Users
interact with this platform-independent format through libraries, with most
high-level analysis languages such as {\sc idl} and
{\color{blue}\href{https://www.python.org/}{\sc python}} able to read variables
from such files directly by name. We provide examples of how to do this in {\sc
python} in Section \ref{sect:python_examples}.  Files can also be queried in
compiled languages such as {\sc c} or {\sc fortran}, once the {\sc hdf5}
libraries are installed. The {\sc hdf5} files can be directly visualised with
an {\sc hdf5} viewer, for example
{\color{blue}\href{https://www.hdfgroup.org/hdf-java-html/hdfview/}{hdfview}}.

\subsection{\hdf\ groups in the snapshots}
\label{sect:fof}
Each snapshot files contains a set of groups. \gadget\ allows for 6 different
particles types (labelled 0-5). Properties of these particles are written in
groups PartType0 to PartType5. In \eagle\ type 0 are gas particles, type 1 are
dark matter particles, type 4 are stellar particles, and type 5 are
supermassive black holes. Types 2 and 3 are not used. We briefly describe the
contents of each group next, see also Table~\ref{table:groups}.

\subsubsection{\texttt{Config}}
The svn subversion revision number of the code that wrote this snapshot, and a
list of all the \gadget\ configuration options set when this code was compiled.
\subsubsection{\texttt{Constants}} Values of physical constants used in the
calculation.

\subsubsection{\texttt{HashTable}}
\label{sect:hash_table}
Particles of each type are distributed across the different files of a snapshot
in such a way that it is easy to retrieve those that are in a simply connected
region - for example all particles within a given distance from a given
location, say the centre of mass of a halo.  This is done by dividing the
computational volume into cubic cells, $2^6$ cells on a side, calculating which
3D cell each particle is in (referred to as the hash key), and sorting
particles based on this hash key. Cells are then distributed over the
individual files that make up a single snapshot. The hash tables allow one to
determine which files need reading to retrieve all particles in a spherical
region around a given centre. We provide a reading routine to do this, and an
example of its use in \ref{sect:python_examples}.  Because particles are
arranged in cubic cells, using the hash tables returns particles in cells -
limiting the list to particles in a spherical region is left to the user. Hash
tables are constructed {\em separately} for each particle type.

\subsubsection{\texttt{Header}}
\label{sect:header}
This contains the standard simulation parameters from \gadget, with some
\eagle\ specific additions. As with standard \gadget, the arrays
\texttt{NumPart\_ThisFile} and \texttt{NumPart\_Total} contain the numbers of
particles of each type (0-5) in the current file, and in all of the files that
constitute the snapshot, respectively. \texttt{MassTable} contains the particle
masses for those particle types that all have the same mass - in this case
\texttt{PartType1}. \texttt{BoxSize} is the linear extent of the simulation
cube. Units of mass and length are the same as for the mass variables of all
particle types (described below), and coordinates, respectively.
\texttt{Omega0} (total matter density in units of the critical density,
$\Omega_{\mathrm{m}}$), \texttt{OmegaLambda} (density parameter corresponding
to the cosmological constant, $\Omega_\Lambda$), \texttt{OmegaBaryon} (mean
baryon density in units of the critical density, $\Omega_{\mathrm{b}}$), and
the \texttt{HubbleParam} ($H_0/(100~{\rm km~s}^{-1}{\rm Mpc}^{-1})\equiv h$ are
taken from \cite{Planck13}. \texttt{ExpansionFactor} is the current value of
the expansion factor $a$, and \texttt{Redshift}=$z\equiv (1/a) - 1$. As in
\gadget, the variable \texttt{Time} is also the expansion factor in these
cosmological runs, it is {\bf not} the age of the Universe. The variable
\texttt{E(z)}$\equiv (\Omega_{\mathrm{m}}/a^3+\Omega_\Lambda)^{1/2}$.

\subsubsection{\texttt{Units}}
Assumed code units of length, time, and mass, and those derived from it, in cgs
(centimetre, grams, and seconds).  Cosmological variables may in addition
depend on powers of $h$ and $a$ as detailed below.  Readers may recognise these
units as Mpc for length, $10^{10}$~M$_\odot$ for mass and km~s$^{-1}$ for
velocity.

\subsubsection{\texttt{Parameters}} A list of the 9 species (chemical elements)
tracked individually in the simulation (H, He, C, Ni, O, Ne, Mg, Si, Fe), their
assumed primordial, and solar abundances, and the assumed metallicity of the
Sun. Note that solar abundances and the metallicity of the Sun are not used in
the code. The radiative cooling and heating interpolation tables used are
described by \cite{Wiersma09a}, these also use Ca and S with ratios provided in
this group. The values of the abundances are collected from the literature and
summarised in Table~1 of \cite{Wiersma09a} - most \eagle\ papers use these
values to convert metal mass fractions into abundances in units of the \lq
solar\rq\ abundance.

\subsubsection{\texttt{RuntimePars}}
This group contains all parameters used by the simulation, from directories for
input and output, over cosmological parameters, to assumed units (these are
written in single precision, which is why the mass unit appears as infinite).
This list also contains the (Plummer equivalent) co-moving and maximum physical
values of the softening length. As per the \gadget\ convention, particle types
0-5 are referred to here as PartType0 = gas, PartType1 = dark matter = \lq
halo\rq, PartType4 = stars = \lq Stars\rq, and PartType5 = black holes = \lq
Bndry\rq.  PartType2 (\lq disk \rq) and PartType3 (\lq bulge \rq) are not used.

\subsubsection{\texttt{PartType0-5}}
\label{sect:part}
Type0 = gas, Type1 = dark matter, Types2 and 3 are not used, Type4 = stars and
Type5=black holes. All particles\footnote{Note that the mass of PartType1 (dark
matter variables) is the same for all particles, and stored in the
\texttt{MassTable} array in the \texttt{Header} group.} have a mass, position,
velocity, and a unique particle identifier (snapshot variables \texttt{Mass},
\texttt{Coordinates}, \texttt{Velocity} and \texttt{ParticleIDs}), but
different types may in addition have a large number of other variables, some of
which are described below. Each variable in the \hdf\ file consists of an array
of numerical values\footnote{The type of variable (float, double, integer) and
rank of the array can be queried in \hdf.}, and 4 attributes that describe the
variable.  Taking as an example the coordinates of a particle (variable
\texttt{Coordinates}), these attributes are
\texttt{CGSConversionFactor}=$3.08\times 10^{24}$,
\texttt{h-scale-exponent}=-1, \texttt{aexp-scale-exponent}=1 and
\texttt{VarDescription}=\lq Co-moving coordinates. Physical position: $r =
ax$ = Coordinates $h^{-1}~a~U_L$ [cm]\rq. 

The variable description is a text string that clarifies what this variable
represents. In the case of the \texttt{Coordinates}, the numerical values
stored are co-moving coordinates, in units of $h^{-1}$~Mpc. The proper position
of a particle is therefore \begin{equation} {\bf r} =
\texttt{Coordinates}\,a^\texttt{aexp-scale-exponent}\,h^\texttt{h-scale-exponent}\,\texttt{CGSConversionFactor}~{\rm
cm}\,.  \label{eq:coordinates} \end{equation} The convention of specifying the
cgs unit, and how the proper variable depends on its co-moving counterpart in
terms of powers of $a$ and $h$, is used for all variables. As another example,
peculiar velocity and particle mass are obtained as

\begin{eqnarray}
{\bf v}&\equiv& a{d{\bf x}\over dt} = \texttt{Velocity}\,a^{1/2}\,h^0\,1\times 10^5~{\rm cm~s}^{-1}\nonumber\\
m &=& \texttt{Mass}\,a^0\,h^{-1}\,1.989\times 10^{43}~{\rm g}\,.	
\end{eqnarray}

\section{Description of all variables}
Tables \ref{table:gas}, \ref{table:dm}, \ref{table:star} and \ref{table:bh}
list descriptions for the particle properties appearing in the snapshot output.
Most time-dependent variables are {\em predicted} to the current snapshot time,
so for example the density variable in the snapshot file is $\rho(t) =
\rho(t_0)+\dot\rho(t_0)\times (t-t_0)$, where $\rho(t_0)$ is the density at
time $t_0$, the last time the density was computed using SPH, and $\dot\rho$ is
an estimate of the rate of change of the density. Because this prediction is
not perfect, computing the SPH density for a particle, given the positions,
smoothing lengths and masses of all other particles in the snapshot, will in
general yield a different value for $\rho$. For most particles, these two
estimates of the density should be close. Note that the SPH smoothing lengths
are also predicted.

Smoothing lengths ($h$) of gas particles are predicted as per the method of
\cite{hopkins2013}, whereby the SPH particle density

\begin{equation}
\label{eq:rho}
\rho_i = \sum_{j=1}^N m_j W_{ij}(h_i),
\end{equation}
	
\noindent where $m_j$ is the mass of each other particle and $W_{ij}(h_i)$ is
the value of the kernel\footnote{\eagle\ uses the ${\cal C}_2$ kernel of
\cite{Wendland1995}.} at that location, yields a proportionality to the
smoothing length of $h_i \propto \rho_i^{-1/3}$, such that the relationship $(4
\pi / 3) h_i^{3} \rho_i = m_i N_{\mathrm{ngb}}$ holds true for a given choice
of $N_{\mathrm{ngb}}$, referred to as the \lq effective neighbour number\rq\
(see \cite{hopkins2013} and Appendix~A1 of \cite{schaye2015} for details).
$N_{\mathrm{ngb}}$ is chosen to be 58 for gas particles. The smoothing lengths
for star and black hole particles are also predicted from the neighbouring gas
particles. However as they are not gas particles themselves, the smoothing
length is now computed ensuring that the relation $(4 \pi / 3) h_i^{3}
\sum_{j=1}^N W_{ij}(h_i) = N_{\mathrm{ngb}}$ holds true for a given choice of
$N_{\mathrm{ngb}}$. For stars $N_{\mathrm{ngb}}$ is chosen to be 48, and 58 for
black holes. 

A short description of all variables is given in the tables below, together
with a reference to an equation clarifying the meaning of the variable taken
from \cite{schaye2015}, see also \cite{mcalpine2016}. We begin by giving some
more information about those variables whose meaning is difficult to convey in
a single sentence.

\subsection{Gas particle variables - \texttt{PartType0}}
\subsubsection{Thermodynamic variables}
\label{sect:thermo}
\eagle\ uses a variety of thermodynamic variables, and it may be important to
realise how they are related, and which version appears in the particle files.
In this {\sc anarchy} version of pressure-entropy SPH, each gas particle $i$
carries its (pseudo) entropy $A_i$. This is the particle variable
\texttt{Entropy}.

The entropy variable appears in the expression for the hydrodynamical
acceleration, calculated from \begin{equation} \frac{{\rm d}\mathbf{v}_i}{{\rm
d}t}|_{\rm hydro}=-\sum_{j=1}^N m_j  \left[
\frac{A_j^{1/\gamma}}{A_i^{1/\gamma}}\frac{\bar{P}_i}{\bar{\rho}_i^2}f_{ij}\nabla_i
W_{ij}(h_i) ~ +
\frac{A_i^{1/\gamma}}{A_j^{1/\gamma}}\frac{\bar{P}_j}{\bar{\rho}_j^2}f_{ji}\nabla_i
W_{ij}(h_j) \right] , \label{eq:mot} \end{equation} where the entropy-weighted
pressure, $\bar{P}$, and density, $\bar{\rho}$, are computed for each particle
from \begin{equation} \bar{P}_i = A_i
\left(\frac{1}{A_i^{1/\gamma}}\sum_{j=1}^N m_j A_j^{1/\gamma}
W_{ij}(h_i)\right)^{\gamma} \equiv A_i \bar{\rho}_i^{\gamma} \,, \label{eq:pS}
\end{equation} $f_{ij}$ are \lq grad-$h$ \rq terms, and $\gamma=5/3$ is the adiabatic index. For
gas in hydrostatic equilibrium, the acceleration as computed from
Eq.~(\ref{eq:mot}) is balanced by gravity. 

Shocks will change $A_i$ at a rate consistent with Eq.~(\ref{eq:mot}).
Radiative cooling and heating, feedback, and the imposed pressure floor may
also change $A_i$, as described next.  The radiative rates depend on the
particle temperature, $T$, and density Eq.~(\ref{eq:rho}), which are computed
for each particle $i$ as

\begin{eqnarray}
\label{eq:u}
u_i &=& {S_i\,\bar{\rho}^{\gamma-1}\over \gamma-1} = {kT_i\over (\gamma-1)\mu_i\,m_{\rm H}},\\
\label{eq:mu}
\mu_i &=& \mu_i(u_i, \rho_i, J(\nu), X, Y, Z),\\
\label{eq:cool}
\rho_i\,{{\rm d}u_i\over {\rm dt}} &=& -\Lambda(T_i)\rho_i^2+{\cal H}(T_i)\,\rho_i\,.
\end{eqnarray}
The conversion from thermal energy per unit mass, $u$, to temperature $T$,
depends on the mean molecular weight (in units of the proton mass, $m_{\rm
H}$), $\mu$. $\mu$ is computed using the interpolation tables described by
\cite{Wiersma09a}, which accounts for the element abundances of the particle,
the thermal energy per unit mass, $u_i$, the density, $\rho_i$, and the
radiation field $J(\nu)$, as is symbolically illustrated by Eq.~(\ref{eq:mu}).
Note that the density $\rho$ in Eq.~(\ref{eq:rho}), Eq.~(\ref{eq:mu}) and
Eq.~(\ref{eq:cool}) is the usual SPH density, which differs from the
entropy-weighted density that appears in Eq.~(\ref{eq:mot}), Eq.~(\ref{eq:pS}),
and Eq.~(\ref{eq:u}).

In the feedback routines, particle temperatures may be increased by a fixed
increment ($\Delta T=10^{7.5}$~K in the case of stellar feedback for the {\sc
reference model}). This is implemented by increasing the entropy of the particle
by an amount computed from Eq.~(\ref{eq:u}). \eagle\ also imposes a pressure
floor and a minimum temperature floor ($T>100$~K). After feedback (but before
cooling), we calculate $T_i$ for every (active) gas particle. We use this to
compute the maximum temperature a particle had throughout its history, as well
as the expansion factor that corresponds to this event.

The radiative equation Eq.(\ref{eq:cool}) calculates the new value of $u$,
evaluating the rate at constant $\rho$. The new value of $u$ is used to update
$S$ and $\dot S\equiv (S(t+dt)-S(t))/{\rm d}t$, where ${\rm d}t$ is the current
time step.

The pressure floor is of the form $p\ge p_{\rm lim}\,(\rho/\rho_{\rm
lim})^{\gamma_{\rm lim}}$,  where $p_{\rm lim}$, $\rho_{\rm lim}$ and
$\gamma_{\rm lim}$ are constants, and is applied if the density is above a
given proper density threshold, $\rho_{\rm lim}$, as well as above an
overdensity threshold, $\Delta_{\rm lim}$. Two thresholds are imposed, which
are expressed in terms of the corresponding temperature thresholds $T_{\rm
lim}$ and hydrogen number density thresholds, $n_{\rm H, lim}$, related by
$n_{\rm H, lim}=X\rho_{\rm lim}/m_{\rm H}$ and $p_{\rm lim}=\rho_{\rm
lim}T_{\rm lim}/(\mu\,m_{\rm H})$, where $\mu=4/(4-3Y)\approx 1.23$ is the mean
molecular weight for neutral gas with the \cite{Planck13} Helium abundance by
mass of $Y=0.248$. For the {\sc reference model}, these values are

\begin{itemize} 

\item A Jeans threshold, for which $T_{\rm lim}=8000$~K, $n_{\rm H,
lim}=0.1$~cm$^{-3}$, $\gamma_{\rm lim}=4/3$, $\Delta_{\rm lim}=10$.

\item A temperature threshold, for which $T_{\rm lim}=8000$~K, $n_{\rm H,
lim}=10^{-5}$~cm$^{-3}$, $\gamma_{\rm lim}=1$, $\Delta_{\rm lim}=10$.

\end{itemize}

In the snapshot files, the variable \texttt{Density}$\equiv \rho$, the variable
\texttt{InternalEnergy}$\equiv u$, and the variable \texttt{Temperature}$\equiv
T$, all of which are predicted to the snapshot time. The variables
\texttt{MaximumTemperature} and \texttt{AExpMaximumTemperature} correspond to
the maximum temperature this gas particle ever had, and the value of the
expansion factor when this occurred, respectively. 

\subsubsection{Abundances} \label{sect:abundances} The abundance group stores
the fraction of a particle's mass in each of the explicitly tracked elements
(H, He, C, N, O, Ne, Mg, Si, Fe). Stellar particles enrich gas particles using
the SPH scheme. During this enrichment step, the particle abundance of an
element, for example C, increases as \begin{equation} \Delta m_{i, {\rm C}} =
\sum_j dm_{j, {\rm C}}\,W_{ij}(h_j)/ \sum_j W_{ij}(h_j)\,, \end{equation} where
$m_{i, {\rm C}}$ is the C mass of gas particle $i$, and the sum is over
(active) star particles. $dm_{j, {\rm C}}$ is the amount of C released by the
three stellar evolutionary channels followed (i.e. AGB stars, type Ia SNe, and
winds from massive stars and their core collapse SNe), over the current time
step of star particle $j$ (i.e {\bf not} using the instantaneous recycling
approximation) .  The variable \texttt{Carbon} in the group
\texttt{ElementAbundance} is then the ratio $m_{i, {\rm C}}/m_i$ of the
particle's mass in C to its total mass. The simulation tracks also a \lq total
metallicity\rq,  - metal mass fraction in all elements more massive then Helium
over total mass - variable \texttt{Metallicity} (note that this includes
contributions from elements that are not tracked individually), the metal mass
fractions from each of the three channels separately (variables
\texttt{MetalMassFracFromAGB}, \texttt{MetalMassFracFromSNIa} and
\texttt{MetalMassFracFromSNII}), as well as the total mass received through
these channels. To study the contribution of type Ia and type II SNe to Fe
enrichment separately, \eagle\ stores the variable
\texttt{IronMassFracFromSNIa} - the ratio of the mass in Fe received through
type Ia SNe only, over the mass of the particle.

\eagle\ also computes \lq SPH-smoothed\rq\ versions of these metal masses by
calculating a \lq metal density\rq\ by summing over gas neighbours. Taking
again C as an example, the metal mass density is \begin{equation} \rho_{i, {\rm
C}} = \sum_j m_{j, {\rm C}}\,W_{ij}(h_i)\,.  \end{equation} The {\em smoothed}
C abundance is then $X_{\rm C}=\rho_{i, {\rm C}}/\rho_i$ and the variable
\texttt{Carbon} in group \texttt{SmoothedElementAbundance} is $X_{\rm C}\,m$
(and similarly for other elements and for other smoothed metallicities). The
motivation for using a smoothed metallicity is explained in \cite{wiersma09b}.

Smoothed abundances are used to calculate the radiative rates in
Eq.~(\ref{eq:cool}), to set the metallicity dependence of feedback,
Eq.(\ref{eq:f(Z,n)}), and to compute stellar evolution. However, the particle
metallicity is used to set the star formation threshold, see below.

\subsubsection{Star formation variables}
\label{sect:sfr}
The star formation rate is compute using the method of \cite{schaye2008}. For a
gas particle $i$, the star formation rate is computed in \eagle\ as
\begin{equation} \dot{m}_{i,\ast} = m_i\, A \left (1~{\rm M}_\odot\,{\rm
pc}^{-2}\right)^{-n} \left ({\gamma \over G} f_{\rm g}
\bar{P}\right)^{(n-1)/2}\,, \label{eq:sflaw} \end{equation} where $A$ and $n$
are constants ($A=1.515\times 10^{-4}$ and $n=1.4$ in the {\sc reference}
model), provided it is eligible for star formation. \eagle\ uses the
metallicity-dependent star formation threshold from \cite{schaye2004}, which is
a fit to the warm, atomic to cold, molecular phase transition, and also
requires the particle to be cold enough (see section 4.3 in \cite{schaye2015}).
The particle's star formation rate is stored in the variable
\texttt{StarFormationRate}.

The slightly misnamed variable \texttt{OnEquationOfState} is a star formation
flag. Its value is 0 if a gas particle has never crossed the star formation
threshold. A positive non-zero value indicates the value of the expansion
factor $a$ when it last became star forming, a negative value indicates the
value of $-a$ when it last failed to meet the star formation threshold.

\subsection{Dark matter particles - \texttt{PartType1}}
This particle group does not include the \texttt{Mass} variable. All dark
matter particles have the same particle mass, found as the second entry in the
array \texttt{MassTable} in the group \texttt{Header}, and with the same units
as all other mass variables. For an example of using this to create an array of
dark matter particle masses in {\sc Python}, see Section
\ref{sect:dm_mass_example}

\subsection{Star particle variables - \texttt{PartType4}}
\label{sect:feedback}
In \eagle, a gas particle may be wholly converted into a star particle. That
star particle inherits all element abundances of its parent gas particle. In
addition, \eagle\ stores the density ($\rho$ - the SPH density) of the gas
particle when it was converted (variable \texttt{BirthDensity}) and the value
of the expansion factor, $a$, when the conversion happened (variable
\texttt{StellarFormationTime}). Note that all these variables are constants
once a gas particle has been converted into a star: they will never change.

The variable \texttt{FeedbackEnergy\_Fraction} is the instantaneous value of
$f_{\rm th}$ from Eq.(7) of \cite{schaye2015}

\begin{equation}
f_{\rm th} = f_{\rm th,min} + \frac{f_{\rm th,max} - f_{\rm th,min}}
{1 + \left (\frac{Z}{0.1\times 0.02}\right )^{n_Z} \left (\frac{n_{\rm H,birth}}{n_{{\rm H},0}}\right )^{-n_n}},
\label{eq:f(Z,n)}
\end{equation} 

\noindent the expectation value of the fraction of the energy released by
type II SNe used to heat gas particles in the stellar feedback implementation.
Note that this is the expectation value of the energy used in the stochastic
implementation of thermal feedback of \cite{DallaVecchia_Schaye2012}. The
normalisation of the (particle metallicity) dependence is 10 per cent of 0.02 -
an approximation to the solar metallicity. The density dependence is calculated
based on the SPH density, $\rho$.

As a star particle ages, \eagle\ evolves the single stellar population with
stellar life-times and evolutionary tracks as described by \cite{wiersma09b}.
As stars evolve, mass and metals are transferred from the star particle to
neighbouring gas particles. The variable \texttt{Mass} is the current particle
mass, whereas \texttt{InitialMass} is the star particle's birth mass.

\subsection{Black hole particle variables - \texttt{PartType5}}
\label{sect:BH}
Black hole (BH) particles are seeded with a given mass in each FoF halo above a
given mass that does not already contain a BH. The expansion factor $a$ of when
the BH was seeded is stored in the snapshot variable
\texttt{BH\_FormationTime}.

BHs may then grow in mass through mergers (with other BHs), and the accretion
of neighbouring gas.  \texttt{BH\_CumNumSeeds} is the total number of seeds the
BH merged with, and \texttt{BH\_MostMassiveProgenitorID} is the
\texttt{ParticleID} of the most massive progenitor of any of the BHs this BH
merged with, and \texttt{BH\_TimeLastMerger} is the value of $a$ when the last
merger occurred.

Following \cite{springel2005b}, \eagle\ uses a subgrid model for BH particles.
The mass of the black hole, $m_{\rm BH}$, which sets its accretion rate, is
allowed to differ from the particle mass, $m$, which is used only for
gravitational calculations. A short summary of the relevant equations, taken
from \cite{schaye2015}, clarifies the meaning of the variables that describe
accretion. The accretion rate is the minimum of the Eddington rate,
\begin{equation} \dot{m}_{\rm Edd} = \frac{4\pi G m_{\rm BH} m_{\rm
H}}{\epsilon_{\rm r} \sigma_{\rm T} c}, \end{equation} ($G$ is Newton's
constant, $\sigma_{\rm T}$ Thomson's cross section and $c$ the speed of light,
and $\epsilon_r$ is the variable \\ \texttt{BlackHoleRadiativeEfficiency} in
the \texttt{RunTimePars} group, $\epsilon_r=0.1$ in the {\sc reference runs})
and

\begin{equation}
\dot{m}_{\rm accr} = \dot{m}_{\rm Bondi} \times \min\left
(C_{\rm visc}^{-1}(c_{\rm s}/V_\phi)^3,1 \right ),
\label{eq:mdotaccr}
\end{equation}

\noindent where $\dot{m}_{\rm Bondi}$ is the Bondi-Hoyle rate for
spherically symmetric accretion, \begin{equation} \dot{m}_{\rm Bondi} =
\frac{4\pi G^2 m_{\rm BH}^2 \bar{\rho}}{(c_{\rm s}^2 + v^2)^{3/2}}.
\end{equation}

\noindent The mass of the BH grows in a time $dt$ by \begin{equation} \Delta
m_{\rm BH} = (1-\epsilon_r)\,\dot m_{\rm accr}\,dt\,. \end{equation} \noindent

\noindent When a BH accretes mass, it stores energy $E$ in a reservoir, which
increases in time step $dt$ by \begin{equation} \Delta E_i =
\epsilon_f\,\epsilon_r\,\dot m_{\rm accr}\,dt\,c^2\,, \end{equation} where
$\epsilon_f=$\texttt{BlackHoleFeedbackFactor}, and is 0.15 in the {\sc
reference} model.

\noindent The sound speed, $c_{\rm s}$, the speed of the gas relative to the BH,
$v\equiv |{\bf v}|$, and the weighted pressure near the BH, $p_{\rm BH}$, are
\begin{eqnarray} c_{\rm s} &=& \left({1\over\bar{\rho}_i}\, \sum_j\,
m_j\,{\gamma\bar{p_j}\over \bar{\rho_j}}\,W_{ij}\right)^{1/2}\,,\nonumber\\
p_{\rm BH} &=& {1\over\bar{\rho}_i}\, \sum_j\,
m_j\,\bar{p_j}\,W_{ij}\,,\nonumber\\ {\bf v}&=&{1\over\bar{\rho}_i}\, \sum_j\,
m_j\, {\bf v}_j \,W_{ij}-{\bf v}_i\,.  \end{eqnarray}

\noindent These sums are over those gas neighbours of the BH that are within its
smoothing length, $h$. Some of the variables that appear in the snapshot,
referring to the last time the BH particle was active, are, \begin{eqnarray} h
&=&
\hbox{\texttt{BH\_SmoothingLength}}=\hbox{\texttt{BH\_AccretionLength}}\nonumber\\
\dot m_{\rm accr} &=& \hbox{\texttt{MH\_Mdot}}\nonumber\\ m_{\rm BH} &=&
\hbox{\texttt{BH\_Mass}}\nonumber\\ \bar{\rho} &=&
\hbox{\texttt{BH\_Density}}\nonumber\\ \bar{p} &=&
\hbox{\texttt{BH\_Pressure}}\nonumber\\ v &=&
\hbox{\texttt{BH\_SurroundGasVel}}\nonumber\\ E &=&
\hbox{\texttt{BH\_Energy\_Reservoir}}\nonumber\\ \end{eqnarray}

\noindent When a BH heats surrounding gas, its energy reservoir is
correspondingly decreased, as described in Section 4.6 of \cite{schaye2015}.

\section{Python code examples}
\label{sect:python_examples}
Below we provide some simple example {\sc Python} scripts that read, process
and display data from the \eagle\ snapshots at $z=0$ ({\sc Snapnum} 28). Each
example is available to download at {\color{blue}
\href{http://icc.dur.ac.uk/Eagle/database.php}{http://icc.dur.ac.uk/Eagle/database.php}}.
These examples assume by default that the snapshot data are located in a folder
named \lq data\rq and in order for these routines to work, the user must have {\sc
Numpy} ({\color{blue}\href{http://www.numpy.org/}{http://www.numpy.org/}}), {\sc
MatPlotLib}
({\color{blue}\href{https://matplotlib.org/}{https://matplotlib.org/}}) and {\sc
AstroPy} ({\color{blue}\href{http://www.astropy.org/}{http://www.astropy.org/}})
installed.

\subsection{Reading datasets}
{\sc read\_dataset} loops over each \hdf\ part (see Section
\ref{sect:eagle_pdata} for how the \hdf\ files are structured) of snapshot 28
and extracts a chosen dataset for all particles of a particular type. It then
converts the data into physical CGS units using information from the dataset's
attributes (see Section \ref{sect:part} for details on unit conversion). For
example, if we wished to extract the physical {\sc Density} for all gas
particles ({\sc PartType} = 0), we would input read\_dataset(0, \lq
Density\rq).  Note that the number of \hdf\ part files ({\sc nfiles}) may be
different depending on the simulation. 

\scriptsize \pythonexternal{python_examples/read_dataset.py} \normalsize

\subsection{Reading the {\sc Header} group} {\sc read\_header} reads
information from the {\sc Header} group (see Section \ref{sect:header} for
details). This example reads and returns the scale factor, Hubble parameter and
the simulation box size (stored in cMpc/h units).

\scriptsize \pythonexternal{python_examples/read_header.py} \normalsize

\subsection{Reading dark matter mass}
\label{sect:dm_mass_example}
As all dark matter particles share the same mass, there exists no {\sc
PartType1/Mass} dataset in the snapshot files.  Instead, the dark matter mass
is stored in the {\sc MassTable} attribute (index 1) in the {\sc Header}
group. Below is an example function that uses this information to create a mass
array for dark matter particles. The array length is determined by the {\sc
NumPart\_Total} attribute in the {\sc Header} and conversion factors are taken
from the {\sc PartType0/Mass} attributes. This is the only dataset that needs
special treatment, all other datasets can be read using the previous example.

\scriptsize \pythonexternal{python_examples/read_dataset_dm_mass.py}
\normalsize

\subsection{Plotting the rotation curve of a galaxy} \label{sect:rotcurve}
Here we use the three functions defined above to plot the rotation curve of the
largest central galaxy ({\sc GroupNumber} 1, {\sc SubGroupNumber} 0) from the
small test volume ({\sc Ref}L0012N0188). We find the centre ({\sc CentreOfMass\_x} =
12.08808994, {\sc CentreOfMass\_y} = 4.47437191, {\sc CentreOfMass\_z}
1.41333473 Mpc) using the public database.

The script loads the {\sc Coordinates} and {\sc Mass} for all gas, dark matter,
star and black hole particles from the selected galaxy using the {\sc
read\_galaxy} function in conjunction with the functions described above and
plots the rotation curve for each component. Note that we must wrap all
coordinates around the centre in order to account for the simulation's
spatial periodicity.

\scriptsize \pythonexternal{python_examples/RotationCurve.py} \normalsize

\subsection{Plotting the temperature--density relation for a galaxy}
\label{sect:phasediag} This example uses a function called {\sc read\_galaxy}
to read the {\sc Temperature}, {\sc Density} and {\sc StarFormationRate} for
each gas particle in the largest central galaxy ({\sc GroupNumber} 1, {\sc
SubGroupNumber} 0) from the small test volume ({\sc Ref}L0012N0188). It then plots
{\sc Temperature} vs {\sc Density} with data points coloured red for a non-zero
{\sc StarFormationRate} and blue otherwise.

\scriptsize \pythonexternal{python_examples/PhaseDiagram.py} \normalsize

\subsection{The {\sc read\_eagle} routine}
In the previous examples we read the entire particle data from the small test
volume ({\sc Ref}L0012N0188) and masked to only particles contained within a
single galaxy. For much larger volumes, particularly the {\sc Ref}L0100N1504
simulation, this becomes increasingly impractical and memory intensive. For
this reason, it is more manageable to read these simulations utilizing the {\sc
HashTable} (see Section \ref{sect:hash_table} for details). 

To aid with this, we provide a reading routine, named {\sc read\_eagle}, that
reads simulation datasets efficiently using the {\sc HashTable}.  {\sc
read\_eagle} is publicly available via a {\sc git} repository, located at
{\color{blue}\href{https://github.com/jchelly/read\_eagle}{https://github.com/jchelly/read\_eagle}}.
The module must first be installed, for instructions on how to do this for {\sc
Python} (instructions for use with {\sc C} and {\sc Fortran} languages are also
available) please refer to the {\sc README} documentation provided within the
repository. 

\subsubsection{{\sc read\_eagle} example}
Below we provide an example that is a repeat of a previous example, where we
create a temperature--density relation for a single galaxy, however this time
using the {\sc read\_eagle} routine. 

First, {\sc read\_eagle} must read information from the {\sc HashTable} and
{\sc Header}. This is done by initializing the {\sc EagleSnapshot} class with
the location of \textit{any} \hdf\ part of the snapshot data (it does not have
to be part 0). You must then select a region of interest. This is a cubic
region outlined by the minimum and maximum extents in the $x$, $y$ and $z$
directions (in cMpc/h units). For our example, we extract a 2 cMpc/h cube that
is centred on the galaxy's centre of potential (taken from the database).
Datasets can then be read in a similar fashion to the examples above that did
not use {\sc read\_eagle}, using the {\sc read\_dataset} routine.

Note there are additional examples within the repository, including {\sc C} and
{\sc Fortran} examples, that explain the further functionality of {\sc
read\_eagle} beyond simply reading datasets (for example reading files in
parallel). Any queries or bugs discovered using this module can be reported via
the repository.

\scriptsize \pythonexternal{python_examples/PhaseDiagram_ReadEagle.py} \normalsize

\newpage
\begin{figure}
\centering
\includegraphics[width=10cm]{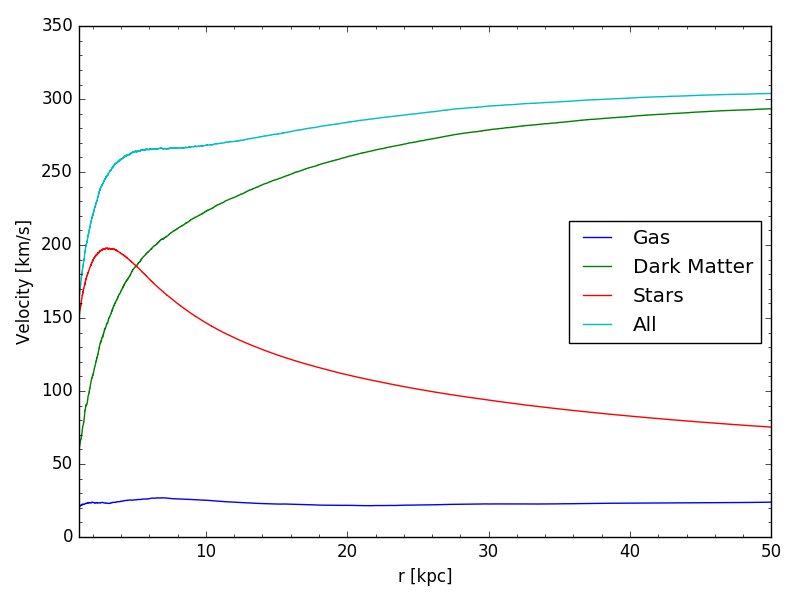}
\caption{Rotation curve produced by the example in \ref{sect:rotcurve}.}
\label{fig:phasediag}
\end{figure}

\begin{figure}
\centering
\includegraphics[width=10cm]{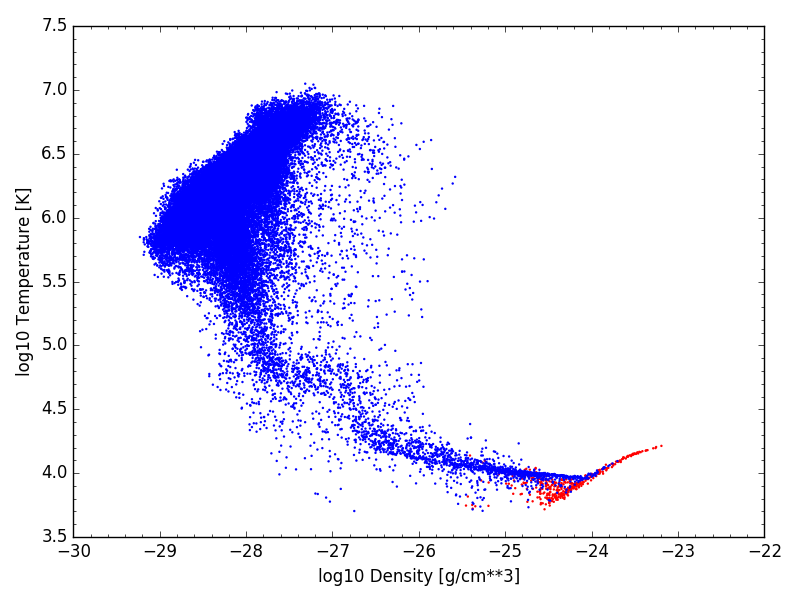}
\caption{Phase diagram produced by the example in \ref{sect:phasediag}.}
\label{fig:rot_curve}
\end{figure}

\newpage

\begin{table}
\label{table:gas}
\caption{Description and equation, where applicable, for each property of gas
(PartType0) particles.}
\begin{center}
\footnotesize
\renewcommand{\arraystretch}{1.5}
\begin{tabular}{>{\ttfamily}p{4cm}p{1.5cm}p{11cm}}
\multicolumn{3}{l}{\large \bf PartType0: Gas} \\
\hline
Field & Equation/Par & Description \\ \hline\hline

AExpMaximumTemperature &
\S\ref{sect:thermo} &
Expansion factor when particle had maximum temperature. \\

Coordinates &
Eq.~(\ref{eq:coordinates}) &
\coordinates \\

Density &
Eq.~(\ref{eq:rho}) &
Co-moving density. \\

ElementAbundance &
\S\ref{sect:abundances} &
Mass of [Carbon, Helium, Hydrogen, Iron, Magnesium, Neon, Nitrogen, Oxygen, Silicon] divided by particle mass. \\

Entropy &
\S\ref{sect:thermo} &
Particle entropy. \\

GroupNumber & \S\ref{sect:fof} & \groupnumber \\

HostHalo\_TVir\_Mass &
- &
Estimate of host FoF group's virial temperature, calculated from the local velocity dispersion. \\

InternalEnergy &
Eq.~(\ref{eq:u}) &
Thermal energy per unit mass. \\

IronMassFracFromSNIa &
\S\ref{sect:abundances} &
Mass of Iron from SNIa divided by particle mass. \\

Mass &
\S\ref{sect:part} &
Particle mass.\\

MaximumTemperature &
S\ref{sect:thermo} &
Maximum temperature ever reached by particle. \\

MetalMassFracFromAGB &
\S\ref{sect:abundances} &
Mass of metals received from AGB divided by particle mass. \\

MetalMassFracFromSNII &
\S\ref{sect:abundances} &
Mass of metals received from SNII divided by particle mass. \\

MetalMassFracFromSNIa &
\S\ref{sect:abundances} &
Mass of metals received from SNIa divided by particle mass. \\

Metallicity &
\S\ref{sect:abundances} & Mass of elements heavier than Helium, including
those not tracked individually, divided by particle mass\\

OnEquationOfState &
\S\ref{sect:sfr} & 0 if particle has never been star-forming, +ve if currently star-forming,
-ve if not currently star-forming. Value indicates scale factor at which it
obtained its current state. Note this does not ensure the gas particle is on
the equation of state for +ve values, as gas particles can yield non-zero star
formation rates  up to 0.5~dex above the equation of state. \\

ParticleIDs &
\S\ref{sect:part} & Unique particle identifier. Index encodes the particles position in the
initial conditions (see Appendix of \cite{schaye2015} for details). \\

SmoothedElementAbundance &
\S\ref{sect:abundances}  & SPH kernel weighted ElementAbundance (see Section 2.2 of \cite{springel2005} for
the description of kernel weighted properties in SPH). \\

SmoothedIronMassFracFromSNIa &
\S\ref{sect:abundances}  & SPH kernel weighted IronMassFracFromSNIa. \\

SmoothedMetallicity &
\S\ref{sect:abundances}  &
SPH kernel weighted Metallicity. \\

SmoothingLength &
\S\ref{sect:part} &
Co-moving SPH smoothing kernel. \\

StarFormationRate &
Eq.~(\ref{eq:sflaw}) &
Instantaneous star formation rate. \\

SubGroupNumber &
\S\ref{sect:fof} & \subgroupnumber \\

Temperature &
\S\ref{sect:thermo} &
Temperature \\

TotalMassFromAGB &
\S\ref{sect:abundances} &
Total mass received from AGB. \\

TotalMassFromSNII &
\S\ref{sect:abundances} &
Total mass received from SNII. \\

TotalMassFromSNIa &
\S\ref{sect:abundances} &
Total mass received from SNIa. \\

Velocity &
S\ref{sect:part} & \velocity \\

\hline
\end{tabular}
\end{center}
\end{table}

\begin{table}
\label{table:dm}
\caption{Description and equation, where applicable, for each property of dark matter
(PartType1) particles.}
\begin{center}
\footnotesize
\renewcommand{\arraystretch}{1.5}
\begin{tabular}{>{\ttfamily}p{4cm}p{1.5cm}p{11cm}}
\multicolumn{3}{l}{\large \bf PartType1: Dark Matter} \\
\hline
Field & Equation/Par & Description \\ \hline\hline

Coordinates &
\S\ref{sect:part} &
\coordinates \\

GroupNumber &
\S\ref{sect:fof} &
\groupnumber \\

ParticleIDs &
\S\ref{sect:fof} & Unique particle identifier. Index encodes the particles position in the
initial conditions (see \cite{schaye2015} for details). \\

SubGroupNumber &
\S\ref{sect:fof} &
\subgroupnumber \\

Velocity &
\S\ref{sect:part} & \velocity \\

\hline
\end{tabular}
\end{center}
\end{table}

\begin{table}
\label{table:star}
\caption{Description and equation, where applicable, for each property of star
(PartType4) particles.}
\begin{center}
\footnotesize
\renewcommand{\arraystretch}{1.5}
\begin{tabular}{>{\ttfamily}p{4cm}p{1.5cm}p{11cm}}
\multicolumn{3}{l}{\large \bf PartType4: Stars} \\
\hline
Field & Equation/Par & Description \\ \hline\hline

AExpMaximumTemperature &
\S\ref{sect:thermo} &
Expansion factor a when particle had highest temperature. \\

BirthDensity &
\S\ref{sect:sfr} & Local gas density when gas particle was converted to star\\

Coordinates &
\S\ref{sect:part} &
Co-moving Coordinates. \\

ElementAbundance &
\S\ref{sect:abundances} &
Mass of [Carbon, Helium, Hydrogen, Iron, Magnesium, Neon, Nitrogen, Oxygen,
Silicon] divided by particle mass. \\

Feedback\_EnergyFraction &
S\ref{sect:feedback} &
Fraction of energy used for thermal feedback, defined such that a value of
unity corresponds to the energy released by the population of Type II SNe (each
producing 10$^{51}$ erg) described by a Chabrier IMF, with SNe progenitors
having initial masses of 6-100 M$_{\odot}$. \\

GroupNumber &
\S\ref{sect:part} &
\groupnumber \\

HostHalo\_TVir &
- &
Halo's virial temperature. \\

HostHalo\_TVir\_Mass &
- &
Estimate of host FoF groups virial temperature, calculated from the local
velocity dispersion. \\

InitialMass &
\S\ref{sect:feedback} &
Mass at formation time. \\

IronMassFracFromSNIa &
- & Mass of Iron from SNIa divided by particle mass. \\

Mass &
\S\ref{sect:feedback} & Current particle mass. \\

MaximumTemperature &
\S\ref{sect:thermo} & Maximum temperature ever reached by particle. \\

MetalMassFracFromAGB &
\S\ref{sect:abundances} &
Mass of metals received from AGB divided by particle mass. \\

MetalMassFracFromSNII &
\S\ref{sect:abundances}  &
Mass of metals received from SNII divided by particle mass. \\

MetalMassFracFromSNIa &
\S\ref{sect:abundances}  &
Mass of metals received from SNIa divided by particle mass. \\

Metallicity &
\S\ref{sect:abundances}  & Mass of elements heavier than Helium, including
those not tracked individually,  divided by particle mass. \\

ParticleIDs &
\S\ref{sect:part} &
Unique particle identifier. ID is inherited from parent gas particle. \\

PreviousStellarEnrichment &
\S\ref{sect:feedback} &
Expansion factor when this star particle last enriched its neighbours. \\

SmoothedElementAbundance &
\S\ref{sect:abundances}  &
SPH kernel weighted ElementAbundance. \\

SmoothedIronMassFracFromSNIa &
\S\ref{sect:abundances}  &
SPH kernel weighted IronMassFracFromSNIa. \\

SmoothedMetallicity &
\S\ref{sect:abundances}  &
SPH kernel weighted Metallicity. \\

SmoothingLength &
\S\ref{sect:part} &
Co-moving SPH smoothing kernel. \\

StellarEnrichmentCounter &
\S\ref{sect:abundances}  &
Number of timesteps since enrichment was last performed by this particle. \\

StellarFormationTime &
\S\ref{sect:feedback} &
Expansion factor when this star particle was born. \\

SubGroupNumber &
\S\ref{sect:part} &
\subgroupnumber \\

TotalMassFromAGB &
\S\ref{sect:abundances} &
Total mass received from AGB. \\

TotalMassFromSNII &
\S\ref{sect:abundances} &
Total mass received from SNII. \\

TotalMassFromSNIa &
\S\ref{sect:abundances} &
Total mass received from SNIa. \\

Velocity &
\S\ref{sect:part}& \velocity \\

\hline
\end{tabular}
\end{center}
\end{table}

\begin{table}
\label{table:bh}
\caption{Description and equation, where applicable, for each property of black hole
(PartType5) particles.}
\begin{center}
\footnotesize
\renewcommand{\arraystretch}{1.5}
\begin{tabular}{>{\ttfamily}p{4cm}p{1.5cm}p{11cm}}
\multicolumn{3}{l}{\large \bf PartType5: Black Holes} \\
\hline
Field & Equation/Par & Description \\ \hline\hline

BH\_CumlAccrMass &
\S\ref{sect:BH} &
Cumulative mass that has been accreted onto this black hole.  \\

BH\_CumlNumSeeds &
\S\ref{sect:BH} &
Cumulative number of black hole seeds swallowed by this black hole. \\

BH\_Density &
\S\ref{sect:BH} &
Co-moving gas density at the location of the black hole. \\

BH\_FormationTime &
\S\ref{sect:BH} &
Scale factor when this black hole was formed. \\

BH\_Mass &
\S\ref{sect:BH} &
Black hole \textit{subgrid} mass (see Appendix of \cite{mcalpine2016} for more details). \\

BH\_Mdot &
\S\ref{sect:BH} &
Instantaneous black hole accretion rate. \\

BH\_MostMassiveProgenitorID &
\S\ref{sect:BH} &
At the time of the last BH-BH merger, this is the ParticleID of the most massive member of the pair. \\

BH\_Pressure &
\S\ref{sect:BH} &
Gas pressure at the location of the black hole. \\

BH\_SoundSpeed &
\S\ref{sect:BH} &
Gas sound speed at the location of the black hole. \\

BH\_SurroundingGasVel &
\S\ref{sect:BH} &
Peculiar velocity of the gas at the location of the black hole. \\

BH\_TimeLastMerger  &
\S\ref{sect:BH} &
Expansion factor when black hole particle last accreted another black hole. 0 if the particle has never accreted another black hole. \\

Coordinates &
\S\ref{sect:part} &
\coordinates \\

GroupNumber &
\S\ref{sect:fof} &
\groupnumber \\

HostHalo\_TVir\_Mass &
- &
Estimate of host FoF group's virial temperature, calculated from the local velocity dispersion. \\

Mass &
\S\ref{sect:part} &
BH \textit{particle} mass. Users should use the black hole subgrid mass ({\sc
BH\_Mass}) for the actual black hole subgrid mass.\\

ParticleIDs &
\S\ref{sect:part} &
Unique particle identifier. ID is inherited from parent gas particle. \\

SmoothingLength &
\S\ref{sect:part} &
Co-moving SPH smoothing kernel. \\

SubGroupNumber &
\S\ref{sect:fof}
&
\subgroupnumber \\

Velocity &
\S\ref{sect:part} & \velocity \\

\hline
\end{tabular}
\end{center}
\end{table}

\section{Acknowledging these data}
This document is \emph{not} intended to serve as a reference for \eagle.  Users
of \eagle\ data are kindly requested to acknowledge and cite the original
sources following the instructions listed in section 4.2 of
\cite{mcalpine2016}, which we repeat here for completeness:

To recognise the effort of the individuals involved in the design and execution
of these simulations, in their post processing and in the construction of the
database, we kindly request the following:

\begin{itemize}
\item Publications making use of the \eagle\ data extracted from the public
database or particle data are kindly requested to cite the original papers
introducing the project \cite{schaye2015,crain2015} as well as the paper
describing the public release of the galaxy data \cite{mcalpine2016}.
\item Publications making use of the database should add the following line in
their acknowledgement section: "We acknowledge the Virgo Consortium for making
their simulation data available. The \eagle\ simulations were performed using
the DiRAC-2 facility at Durham, managed by the ICC, and the PRACE facility
Curie based in France at TGCC, CEA, Bruy\`eresle-Ch\^atel.".
\item Furthermore, publications referring to specific aspects of the subgrid
models, hydrodynamics solver, or post-processing steps (such as the
construction of images or photometric quantities, and the construction of
merger trees), are kindly requested to not only cite the above papers, but also
the original papers describing these aspects. The appropriate references can be
found in section 2 of this paper and in \cite{schaye2015}.
\end{itemize}

\newpage
\bibliographystyle{unsrt}
\bibliography{ParticleRelease}

\end{document}